\documentclass[sigconf]{acmart}

\AtBeginDocument{%
  \providecommand\BibTeX{{%
    \normalfont B\kern-0.5em{\scshape i\kern-0.25em b}\kern-0.8em\TeX}}}

\setcopyright{acmlicensed}
\copyrightyear{2025}
\acmYear{2025}
\acmDOI{XXXXXXX.XXXXXXX}

\acmBooktitle{Proceedings of the CHI 2026 Workshop: Ethics at the Front-End: Responsible User-Facing Design for AI Systems (CHI '26), April 13--17, 2026, Barcelona, Spain}

\usepackage{multirow}
\usepackage{enumitem}
\usepackage{longtable}
\usepackage{csquotes}
\usepackage{colortbl}
\usepackage{float}
\usepackage{placeins}
\usepackage{indentfirst}
\usepackage{hyperref}
\usepackage{makecell}
\usepackage{xcolor}
\usepackage{soul}
\usepackage{listings}
\usepackage{textcomp}
\usepackage{tabularx}
\usepackage{diagbox}
\usepackage{pifont} 
\usepackage[table]{xcolor}
\usepackage{booktabs}
\usepackage[normalem]{ulem}

\begin{document}

\title[{Browser-Level Overlay for User-Controlled PII Management in AI Interactions}]{PII Shield: A Browser-Level Overlay for User-Controlled Personal Identifiable Information (PII) Management in AI Interactions}

\author{Maximilian Holschneider}
\email{mpholsch@media.mit.edu}
\authornote{Authors contributed equally to this work.}
\affiliation{%
  \institution{MIT Media Lab}
  \city{Cambridge}
  \state{Massachusetts}
  \country{United States of America}
}

\author{Saetbyeol LeeYouk}
\email{sbleeyuk@mit.edu}
\authornotemark[1]
\affiliation{%
  \institution{MIT Media Lab}
  \city{Cambridge}
  \state{Massachusetts}
  \country{United States of America}
}

\renewcommand{\shortauthors}{Holschneider and LeeYouk et al.}

\begin{teaserfigure}
    \centering
    \includegraphics[width=\linewidth]{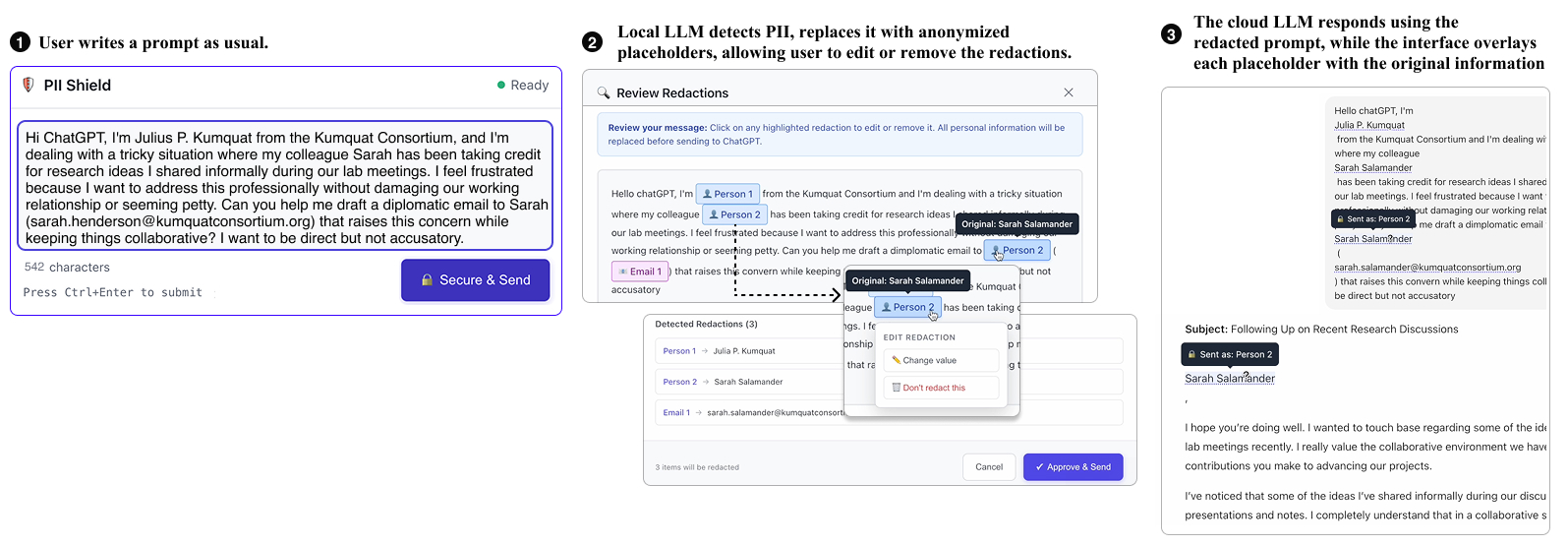}
    \caption{Illustration of how \textit{PII Shield} protects users’ personal information through a browser-level extension overlay. It shows the \textit{PII Redaction} feature, where the local LLM detects personal entities and replaces them with anonymized placeholders before the prompt is sent to the cloud LLM.}
    \label{fig:overview}
\end{teaserfigure}

\begin{abstract}
AI chatbots have quietly become the world's most popular therapists, coaches, and confidants. Users of cloud-based LLM services are increasingly shifting from simple queries like idea generation and poem writing, to deeply personal interactions. As Large Language Models increasingly assume the role of our confessors, we are witnessing a massive, unregulated transfer of sensitive personal identifiable information (PII) to powerful tech companies with opaque privacy practices. While the enterprise sector has made great strides in addressing data leakage concerns through sophisticated guardrails and PII redaction pipelines, these powerful tools have functionally remained inaccessible for the average user due to their technical complexity. This results in a dangerous trade off for individual users. In order to receive the therapeutic or productivity benefits of AI, users need to abandon any agency they might otherwise have over their data, often without a clear mental model of what is being shared, and how it might be used for advertising later on. This work addresses this interaction gap, applying the redaction pipelines of enterprise-grade redaction into an intuitive, first-of-its-kind, consumer-facing, and free experience. Specifically, this work introduces a scalable, browser-based intervention designed to help align user behavior with their privacy preferences during web-based AI interactions. Our system introduces two key mechanisms: local entity anonymization to prevent data leakage, and ‘smokescreens’: autonomous agent activity to disrupt third-party profiling. An open-source implementation is accessible at the following GitHub repository: \url{https://github.com/SBleeyouk/PII_Shield.git}
\end{abstract}



\received{PUT DATE}
\received[revised]{DATE}
\received[accepted]{DATE}

\maketitle

\section{Introduction}

From therapeutic venting to financial planning, Large Language Models (LLMs) have become the de facto companions for our deeply personal tasks. However, this rapidly growing reliance often outpaces our cultural understanding of its security implications, particularly privacy gaps. 

Users are increasingly forced to choose between the utility of AI to help solve knowledge work or personal tasks, and the security of and agency over their personal data. Enterprise environments have made significant strides in recent months in managing data leakage to third party providers through a mixture of guardrails and redaction pipelines, however due to technical and platform gaps, these protections remain functionally inaccessible to the general public.

This work introduces a scalable, browser-based intervention designed to help align user behavior with their privacy preferences during web-based AI interactions. Our system introduces two key mechanisms: local entity anonymization to prevent data leakage, and ‘smokescreens’: autonomous agent activity to disrupt third-party profiling. Based on our system, we propose user study and participatory design workshop as a future work, to investigate whether reducing the barrier to entry for privacy tools can effectively shift users from passive data surrender to active, informed agency. We discuss the implications of these findings for the design of digital environments, arguing that effective personal privacy protection requires interventions that operate at the speed and place of user interaction, rather than relying on post-hoc compliance or high-effort user vigilance.
\section{Related Works}

\subsubsection{\textbf{The Interface as a Locus of Control}}
While AI ethics often scrutinizes backend models, the immediate ethical failures are often most prevalent in the front-end, where interface design significantly shapes user behavior \cite{ibrahim2024characterizingmodelingharmsinteractions}. Current LLM interfaces prioritize friction-free interactions, working to create information architectures where data surrender is the path of least resistance \cite{Allison}. By sidestepping the obscure downstream consequences of data sharing in favor of immediate usefulness, these interfaces function as \textit{dark patterns} \cite{Gunawan}, manipulating users into sharing information that if done more consciously they might not otherwise choose to do so.

Research in behavioral privacy suggests that when users are faced with this trade-off between AI advice (e.g. therapy or financial planning), and abstract privacy risks, the cognitive load required to self-censor trumps the want to keep personal information private \cite{Zhang}. Without explicit privacy choices, the users are subtly guided into sharing content they would otherwise not want to. This is especially dangerous in the context of micro-targeting, as relevant chat context has been found to be useful in belief change \cite{hackenburg2025leverspoliticalpersuasionconversational}.

\subsubsection{\textbf{The Accessibility Gap in the Public Relation Infrastructure}}
As mentioned earlier, the technical challenge of sanitizing sensitive data has made great strides in recent months within enterprise environments, enabling careful control of data leakage from large organizations \cite{nvidia_nemo_guardrails, MsPresidio}. As of yet however, these sophisticated protections have not been translated into accessible interventions for the general public.

In this work, we propose that ethical frontend design includes elements which disrupt the current flow of user data. We believe that while regulation of backend models should be a large concern for regulators and government, for helping individuals reclaim agency, the most useful point of attack sits at the user interface level.

\section{PII Shield: Design Concept}
We define PII as two categories: (1) Direct identifiers, such as names, institutional affiliations, or phone numbers that are explicitly recognizable; and (2) Contextual personal information, such as medical histories, personal finance, or mental-health details that users may provide to obtain personalized assistance but that still enable downstream identification.
From this, we established three design goals (DG): 

\noindent\textbf{DG1.} Preserve user agency by enabling control of unwanted PII leakage.

\noindent\textbf{DG2.} Protect users’ sensitive information while preserving sufficient contextual detail to maintain personalized interactions.

\noindent\textbf{DG3.} Ensure seamless LLM interaction by running PII-protection processes unobtrusively in the background, without introducing additional user burden.

To address these goals, we propose \textit{PII Shield}, a browser-extension interface that provides two complementary controls. The extension overlays existing AI interfaces and functions across platforms. Below, we outline the interactions supported for each PII type.

\subsubsection{\textbf{PII Redaction}} \textit{PII Shield} supports PII redaction for both prompts and file attachments. When users click \textit{Secure Personal Information}, an entity-recognition module replaces detected PII with generalized placeholders (e.g., name → Person A, institution → School A). Redacted text is visually overlaid so the user can still view the original locally. For information not captured by automated detection, users may also manually select additional segments to redact, optionally providing a substitute term or leaving it blank.
The cloud LLM receives only the secured prompt, while the user interface preserves the original version. For model outputs, \textit{PII Shield} displays responses using the redacted placeholders but overlays each placeholder with the corresponding original information in a popup. Users can close or reveal these popups. This ensures that users retain fine-grained control over PII disclosure while minimizing disruptions to the LLM’s performance or the fluidity of the AI interaction.

For uploaded files, the same mechanism detects and masks PII, overlaying redactions in the preview and sending only the redacted version to the LLM. Users can drag to manually redact additional regions. To accommodate diverse file-attachment scenarios, such as providing reference documents or requesting edited versions, the LLM returns an editable format (e.g., .txt for .rtf). Upon receiving the model’s output, \textit{PII Shield} remaps the redacted placeholders to the original PII and renders a final, fully restored version for the user. This ensures that LLM processing occurs on protected content while the user continues to interact with the complete document.

\subsubsection{\textbf{Smokescreens}} 
To achieve DG2, the system introduces a Smokescreen feature when user-provided context may indirectly reveal identity. A local LLM generates a semantically aligned but obfuscated surrogate description, which is injected into the system prompt before cloud inference. This preserves conversational relevance while masking identifiable cues. For instance, a query such as \textit{“I feel exhausted and not willing to do anything”} is reformulated as \textit{“My friend Kevin reports general fatigue and low motivation. Please generate advice directed to Kevin.”} This surrogate is used only internally, ensuring the user continues to receive personalized guidance without exposing contextual information to the cloud model. The mechanism reduces the potential for downstream misuse, such as behavioral profiling, targeted advertising, or inference of mental, financial, or medical conditions.
\section{Future User Study}
As future work, we aim to conduct an expanded user study to examine how individuals perceive privacy risks in cloud-based LLMs and how these perceptions vary across technical backgrounds and usage contexts. Using \textit{PII Shield} as an exploratory probe, the preliminary study will include: (1) an onboarding interview about LLM usage and privacy concerns, (2) hands-on interaction with the system, and (3) a brief follow-up interview on benefits and limitations (see Appendix~\ref{appendix:userstudy}). This phase aims to surface users’ primary PII concerns, their willingness to trade control for personalization, and the usability of client-side protections.

Building on these insights, we will conduct a participatory design workshop involving multiple user groups with varying levels of technical knowledge and LLM experience (e.g., software engineers, AI researchers, creative practitioners, and everyday non-technical users). The workshop will probe how different communities interpret the risks and boundaries of PII disclosure, how they negotiate trade-offs between personalization and privacy, and what kinds of controls they deem meaningful, usable, or burdensome in everyday interaction. Participants will collaboratively critique the current \textit{PII Shield} design, surface unmet needs, and propose alternative interaction patterns or redaction logics. Drawing on methods from community computing~\cite{Karrie2019Algorithms}, the outputs from the workshop will guide the co-design of \textit{PII Shield v2} as an open-source, community-driven privacy layer, ensuring that control over PII remains grounded in user agency rather than platform-level policy.
\section{Conclusion}
We introduced \textit{PII Shield}, a frontend security framework that enables users to control the disclosure of PII during AI conversation. It offers a practical alternative to current cloud-centric privacy protections, preserving conversational utility while limiting unintended data exposure.

Looking ahead, several research directions remain, particularly in applying these ideas to the practice of \textit{vibe coding} knowledge work for individual users. With the recent emergence of desktop agents such as Anthropic's Cowork \cite{anthropic2026cowork}, we see the beginnings of a shift in non-technical individuals increasingly relying on AI tools that gain comprehensive access to one's entire desktop environment. In these high-stakes contexts, where an agent can observe everything that's on a user's machine, the boundary between useful context and unwarranted surveillance becomes murkier. We believe the principles of meaningful, frontend-mediated PII control as shown in our system are especially well-suited for these desktop-level interactions, allowing users to both harness the power of AI for knowledge work without blindly surrendering their digital privacy. 

More broadly, these questions gain urgency as AI platforms move toward integrated, behavior-shaping advertising models. Unlike traditional browser ads, these systems can draw on conversational history and inferred mental, financial, or medical states to target users in highly granular ways. In long-term therapy or other extended use cases, users may need to revisit sensitive disclosures made across sessions, raising additional considerations around data persistence and retrieval. Currently, most AI interactions remain context-window oriented, making this less immediate a concern; moreover, we are optimistic that privacy-based design solutions centered around leveraging local LLMs will continue to keep pace with cloud-based alternatives, as they can cover comparable long-term functionality while keeping sensitive data largely on-device. Designing ethical AI front-ends will therefore require rethinking user agency, transparency, and control to safeguard autonomy in everyday AI use.

\bibliographystyle{ACM-Reference-Format}
\bibliography{References}
\section{Appendix}
\appendix
\section{User Study Protocol}
\label{appendix:userstudy}

\subsection{Onboarding Questionnaire}

\begin{enumerate}
    \item \textbf{How frequently do you use cloud-based LLM systems?}
    \begin{itemize}
        \item Every day
        \item 2--3 times a week
        \item Once a week
        \item Once a month
        \item Rarely / not frequently
    \end{itemize}

    \item \textbf{For what types of tasks do you most frequently use cloud-based LLM systems? (Select all that apply.)} The choices will center around{}: 
    \begin{itemize}
        \item Programming (e.g., “vibe coding,” debugging)
        \item Revising text (e.g., essays, reports)
        \item Revising emails or other personal messages
        \item Mental health or therapeutic support
        \item Medical information or advice
        \item Financial information or advice
        \item Other: \rule{4cm}{0.15mm}
    \end{itemize}

    \item \textbf{When using cloud-based LLMs, do you ever feel concerned about your personal information (e.g., name, institution, medical or financial information) being shared or stored?}
    \begin{itemize}
        \item Yes
        \item No
        \item Not sure
    \end{itemize}

    \item \textbf{What types of personal information are you most concerned about? (Select all that apply.)}
    \begin{itemize}
        \item Name
        \item Affiliation / institution
        \item Location
        \item Government-issued ID
        \item Directory and file access (e.g., via “vibe coding” or file-level tools)
        \item Medical information
        \item Mental health / emotional state
        \item Financial information
        \item Travel plans or destination
        \item Other: \rule{4cm}{0.15mm}
    \end{itemize}

    \item \textbf{Do you currently have your own strategies for protecting your PII when using cloud-based LLMs? If yes, please describe briefly.}
\end{enumerate}

\subsection{Feature Exploration and Follow-up Interview}

After the participant interacts with \emph{PII Shield}, we conduct a short semi-structured interview:

\begin{enumerate}
    \item \textbf{Perceived Benefits (Pros)} \\
    What aspects of \emph{PII Shield} did you find most helpful or reassuring, and why?

    \item \textbf{Perceived Limitations (Cons)} \\
    What aspects of \emph{PII Shield} did you find confusing, inconvenient, or concerning?

    \item \textbf{Adoption Intent} \\
    If this feature were publicly available (e.g., as a browser extension), would you use it in your everyday LLM interactions? Why or why not?

    \item \textbf{Suggestions for Improvement} \\
    Is there anything you would change or add to make \emph{PII Shield} more useful or trustworthy for you?
\end{enumerate}

\end{document}